\documentclass[final]{svjour2}
\usepackage{graphicx}
\usepackage{amssymb}
\usepackage{amsmath}
\usepackage{mathptmx}
\usepackage[dvipsnames]{color}
\usepackage[square,numbers]{natbib}
\usepackage[nodisplayskipstretch]{setspace}
\usepackage{bm}
\makeatletter
\journalname{Journal of Low Temperature Physics}

\DeclareMathAlphabet      {\mathbfit}{OML}{cmm}{b}{it}
\renewcommand{\vec}[1]{\ensuremath{\mathbfit{#1}}} 

\newcommand{\mat}[1]{\ensuremath{\mathbf #1}}   
\newcommand{\inv}[1]{\ensuremath{{\mathbf #1}^{-1}}}   

\newcommand{\be}{\begin{equation}}
\newcommand{\ee}{\end{equation}}
\newcommand{\ba}{\begin{eqnarray}}
\newcommand{\ea}{\end{eqnarray}}

\begin{document}

\title{The Practice of Pulse Processing}

\author{
J.~W.~Fowler, 
B.~K.~Alpert,
W.~B.~Doriese,
Y.-I.~Joe,
G.~C.~O'Neil,
J.~N.~Ullom,
D.~S.~Swetz
}

\institute{National Institute of Standards and Technology, 325
  Broadway, Boulder, CO 80305, USA \\
\emph{Official contribution of NIST, not subject to copyright in the United States.}
}

\date{\vspace{-15mm}\\ November 11, 2015}

\maketitle

\vspace{-6mm}

\begin{abstract}
The analysis of data from x-ray microcalorimeters requires great care; their excellent intrinsic energy resolution cannot usually be achieved in practice without a statistically near-optimal pulse analysis \emph{and} corrections for important systematic errors. We describe the essential parts of a pulse-analysis pipeline for data from x-ray microcalorimeters, including steps taken to reduce systematic gain variation and the unwelcome dependence of filtered pulse heights on the exact pulse-arrival time.  We find these steps collectively to be essential tools for getting the best results from a microcalorimeter-based x-ray spectrometer.

\end{abstract}

\newcommand{\sectionC}[1]{
\vspace{-4mm}
\section{#1}
\vspace{-2mm}}
\newcommand{\subsectionC}[1]{
\vspace{-5mm}
\subsection{#1}
\vspace{-3mm}}

\sectionC{The pulse processing problem}

\begin{figure}[ht]
\includegraphics[width=\textwidth]{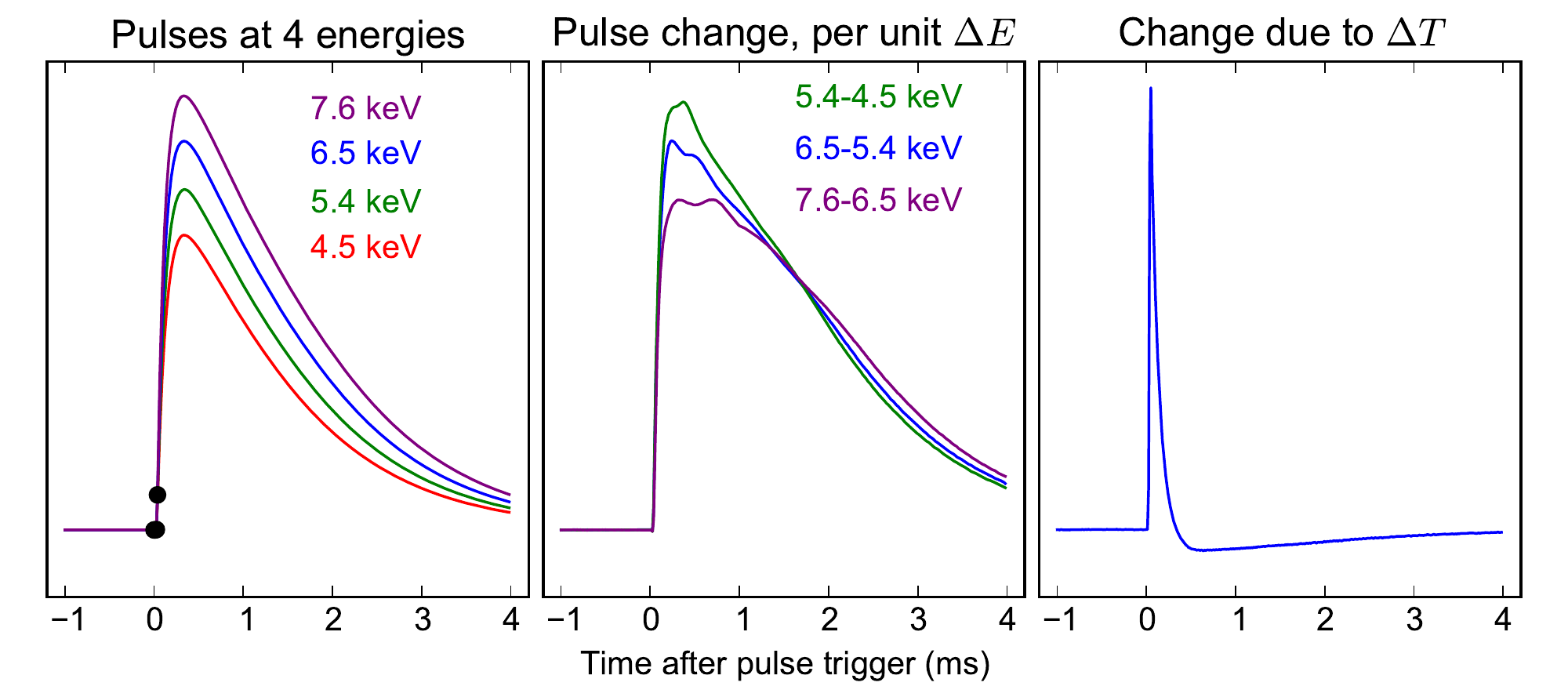}
\caption{ \label{fig:pulses}
Representative TES current pulses and important nonlinear effects. \emph{Left:} Records of $N=390$ samples (5\,ms) for pulses at a range of fluorescence energies.
The trigger decision is made on the four samples indicated by the circles (the first three circles overlap).
\emph{Center:} The difference between the average pulses at a few selected energies, divided by their energy difference. Nonlinearity of the sensors renders these different from one another and from the pulse shape.
\emph{Right:} The average difference at 5.9\,keV between the means of the earliest-arriving and latest-arriving pulses, relative to the trigger time. This shape is similar (but not identical) to the successive-differences of the average pulse. (Color figure online.)
}
\end{figure}

Calorimeters based on the superconducting transition-edge sensor (TES) have demonstrated resolving power $E/\delta E$ above 
2000 at 1.5\,keV \cite{Lee:2015}, 6 keV \cite{Fowler:2015}, 100 keV \cite{Bennett:2012}, and 5 MeV \cite{Hoover:2015}. X-ray 
 data must be treated with great care to reach this level of precision. 
The x-ray photons thermalize in an absorber, creating a \emph{pulse} in the TES bias current (Fig.~\ref{fig:pulses}), whose size indicates the photon energy.
 The statistical uncertainty on the size of each pulse 
must be minimized through statistically optimal filtering, while systematic errors must be kept to an equally low 
level.  

In this report, we describe the procedures that we use for data from spectrometers over a wide range of energies and 
applications.
The figures draw their example data from TESs measuring 4\,keV to 7\,keV x-rays and designed to saturate at 12\,keV\@. 
This survey of data analysis problems and solutions should apply to single-photon energy measurements made with 
TESs over a wide energy range and should be relevant to microcalorimeter data more generally.

To reach the theoretical limits of resolution, we must employ statistically optimal filtering (\S\ref{sec:opt_filt}) to estimate pulse 
heights. Many challenges frustrate our ability to reach these limits, such as slow variations in the overall system gain. This gain drift can be largely corrected because the quiescent, or ``baseline,'' level of the TES readout is found to 
track it well (\S\ref{sec:gain_drift}). 
Also, the estimated pulse amplitude can depend on the exact photon arrival time relative to the data sampling clock (\S
\ref{sec:arrival_time}). 


\sectionC{Optimal filtering for pulse-height estimation} \label{sec:opt_filt} 

We assume as a starting point a ``clean'' data set containing only valid, single-pulse records. We choose this set by 
summarizing each pulse record with a few quantities such as the pretrigger mean and rms deviation, peak time, and the largest 
positive slope found after the peak. These quantities provide a fairly clear picture of pulse quality. They allow us to consider 
only pulses isolated in the record and for which the sensor is in its quiescent state when a pulse arrives.
 
The first step in analyzing the clean microcalorimeter pulses is the estimation of pulse height, generally called 
\emph{optimal filtering}. It consists of a linear operation on the sampled and digitized sensor readings.  The filtering can be performed as a continuous convolution of a fixed-length filter with the data samples; pulse 
heights are then proportional to local maxima in the convolution. In practice, the pulse signal-to-noise ratio is high enough 
that we easily identify pulse arrival times (``trigger'') during the data acquisition phase. We store for later processing a data record 
of $N$ samples---typically hundreds to thousands---omitting those samples long before or after the pulse that contain little relevant 
information. This approach lets us compute and adjust the filter later, in order to select the best balance between systematic 
and statistical errors for a given measurement.  

The linear optimal filter is statistically optimal if and only if the following assumptions about the sensor signal hold:
\begin{enumerate}
\item Pulses are transient departures from a strictly constant baseline level;
\item Regardless of the x-ray energy deposited in the sensor, pulses always have the same shape and differ only in  scale; \label{item:linear}
\item Pulses are separated in time sufficiently that no pulse affects the measurement of another;
\item The noise is additive and stationary (independent of the evolving pulse); and
\item The noise on the $N$ samples is distributed as a multivariate Gaussian.
\end{enumerate} 
None of these assumptions is strictly true (e.g., see Fig.~\ref{fig:pulses} for violations of \#\ref{item:linear}), yet optimal filtering  is a very successful linear framework for pulse height estimation.

To construct the filter, we require a model of the pulse shape and of the noise. The optimal filter is the linear weighting of the $N$ data samples that gives the minimum expected variance at a fixed normalization.
Noise can be characterized either by its power spectral density or by its autocorrelation function. 
In the first case, we would construct the filter using discrete Fourier transforms (DFTs) \cite{Szymkowiak:1993}. In the second, we would use only the noise autocorrelation function \cite{Fixsen:2002}.
We never use the DFT approach in spite of its computational advantage, as there is a small signal-to-noise cost  for the DFT's implicit and incorrect assumption that signal and noise are periodic of period $N$ \cite{Alpert:2012}.

Filters can be constructed subject to one or more linear constraints, to ensure that the estimated pulse height is strictly insensitive to the addition of one or more terms. Nuisance terms might include a constant baseline offset, or a decaying exponential representing residual energy left from previous pulses.
Alpert et al. \cite{Alpert:2013} derive the constrained optimal filter by minimizing the expected variance subject to $N_\mathrm{c}$ constraints, while Fowler et al. \cite{Fowler:2015} derive the same result from a maximum-likelihood fit for a linear model with $1+N_\mathrm{c}$ components.
We quote the result here.

In our standard pulse analysis, we employ a two-component model, consisting of a single pulse shape (having zero baseline level) plus a constant.  The resulting estimation of pulse amplitude is the first component of vector $\vec{\hat{p}}$ in
\be \label{eq:opt_filt}
\vec{\hat{p}} = [(\mat{M}^T \inv{R} \mat{M})^{-1} \mat{M}^T\inv{R}] \vec{d},
\ee
where $\mat{M}$ is the $N\times2$ matrix with the first column containing the pulse model and the second containing all ones; $\mat{R}$ is the noise autocorrelation matrix, an $N\times N$ symmetric Toeplitz matrix whose first row is the noise autocorrelation estimate; and \vec{d} is the column vector of length $N$ containing the measured data. Once the modeling and noise characterization are complete, the $2\times N$ matrix in brackets is fixed and can be pre-computed; we call its first row the ``optimal filter.'' Filtering (or fitting) the data means performing one inner-product operation on each data record to compute pulse heights ($\hat{p}_1$).


\sectionC{Systematic errors}

Given the assumptions of the previous section, the optimal filter promises the best possible estimate of  pulse heights, in the sense of a minimum-variance unbiased estimator. In this section, we consider the two most prominent systematic errors that arise from failures of the linear model.

\subsectionC{Gain drift and spectral entropy as a tool for error correction} \label{sec:gain_drift}

The gain of our TES spectrometers varies over time. Although it falls typically by only one part per thousand over many hours, this change can still degrade the achievable energy resolution. One very helpful fact---empirically observed if not fully understood---is that the reduced gain correlates strongly with a slight increase in the baseline level. A multiplicative correction is routinely applied to all data; the correction is linear in the baseline level:
\begin{equation} \label{eq:drift}
p'_j = p_j[1+\alpha(B_j-B_0)].
\end{equation}
Here $p_j$ and $p'_j$ are the plain and corrected optimal filtered pulse height estimates for pulse $j$; $B_j$ is the baseline level for that pulse (estimated as the mean of all pre-trigger samples); $B_0$ is the median baseline estimate for all clean pulses from that TES; and the one free parameter (per sensor) is the ``gain slope'' $\alpha$, typically of order $10^{-4}$. This simple correction is usually needed in observations lasting at least one hour, and it is adequate even for observations of ten or more hours.

\begin{figure}[ht]
\includegraphics[width=\textwidth]{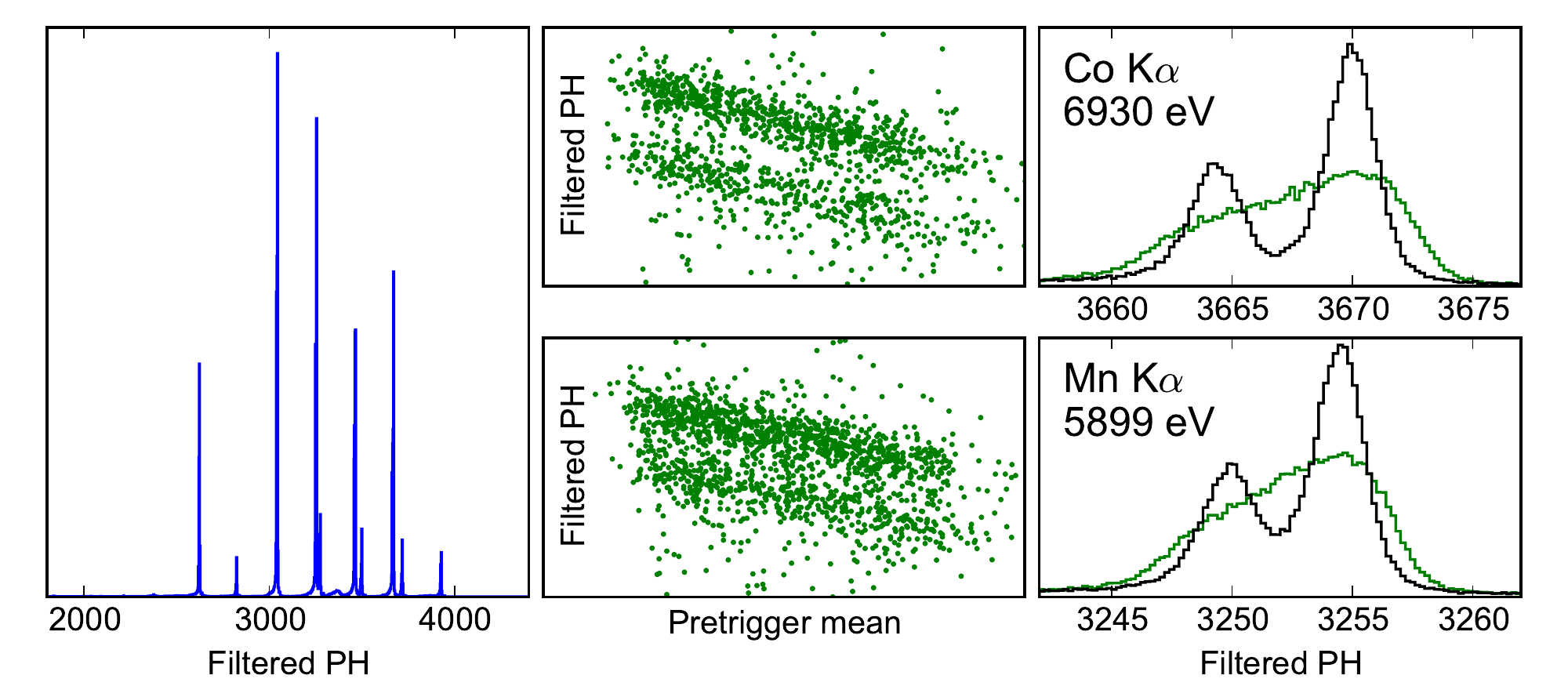}
\caption{ \label{fig:drift_correction}
The gain drift and its correction. \emph{Left:} Filtered pulse heights for the fluorescence of a  sample containing five transition metals. 
\emph{Center:} The anti-correlation between filtered pulse height and the pretrigger mean (baseline) level at the two-peaked cobalt (\emph{top}) and manganese (\emph{bottom}) K$\alpha$ lines.
\emph{Right:} The K$\alpha$ line shapes before (\emph{green/gray}) and after (\emph{black}) application of 
the entropy-minimizing correction for slow drifts in the gain (Eq.~\ref{eq:drift}). After the correction, the FWHM Gaussian resolution is 4.4\,eV and 4.0\,eV at 6.9\,keV and 5.9\,keV.  (Color figure online.)
}
\end{figure}


We choose the gain slope $\alpha$ in Eq.~\ref{eq:drift} considering the entire energy spectrum, rather than a single line.
The training data can be either the entire data set from one sensor or a large subset of it, provided the subset spans the observed range of baseline levels.\footnote{$10^4$ or $10^5$ pulses should easily suffice for selecting $\alpha$.} We are guided by the principle that the ``right'' value of $\alpha$ can be recognized because it leads to the sharpest possible spectrum, with high contrast between any narrow lines and the continuum.\footnote{Naturally, this principle is easier to use when some or all data appear in a few narrow lines.}  This sharp-spectrum notion can be quantified by the Shannon entropy of the observed energy distribution: a low-entropy spectrum has sharper features than a high-entropy spectrum. If the training data are binned into a normalized pulse-height histogram with $N_\mathrm{b}$ bins\footnote{The bin size should be somewhat smaller than the resolution.}, where bin $i$ contains a fraction $f_i$ of all pulses, then the spectral entropy is estimated as:
\be
H = -\sum_{i=1}^{N_\mathrm{b}}\ f_i \log_2\,f_i
\ee
in units of bits per pulse.  We select the correction $\alpha$ that minimizes the spectral entropy $H(\alpha)$, a simple scalar minimization problem. Fig.\,\ref{fig:drift_correction} shows the scale of the problem and the results of an entropy minimization for a single TES sensor.

\subsectionC{Arrival time} \label{sec:arrival_time}

The pulse heights produced by optimal filtering are subject to a large and unwelcome dependence on
the exact arrival time of a photon at the microcalorimeter, relative to the regular sampling clock (typically 5\,$\mu$s to 10\,$\mu$s in our measurements). If the arrival time merely caused a shift in the sampling times of a smooth underlying model curve, then the effect could be large yet conceptually simple. Unfortunately, the limited dynamic range of our active-feedback SQUID readout system can lead to large nonlinearities in the first several samples of a pulse's rising edge and to complicated biases. These biases can degrade the energy resolution in measurements whenever the current through the TES changes during one sample period by an amount that exceeds the SQUID amplifier's linear range. This effect should not be unique to time-division multiplexing, but should occur whenever a readout system has nonlinear response on the pulse leading edge.

\begin{figure}[ht]
\includegraphics[width=\textwidth]{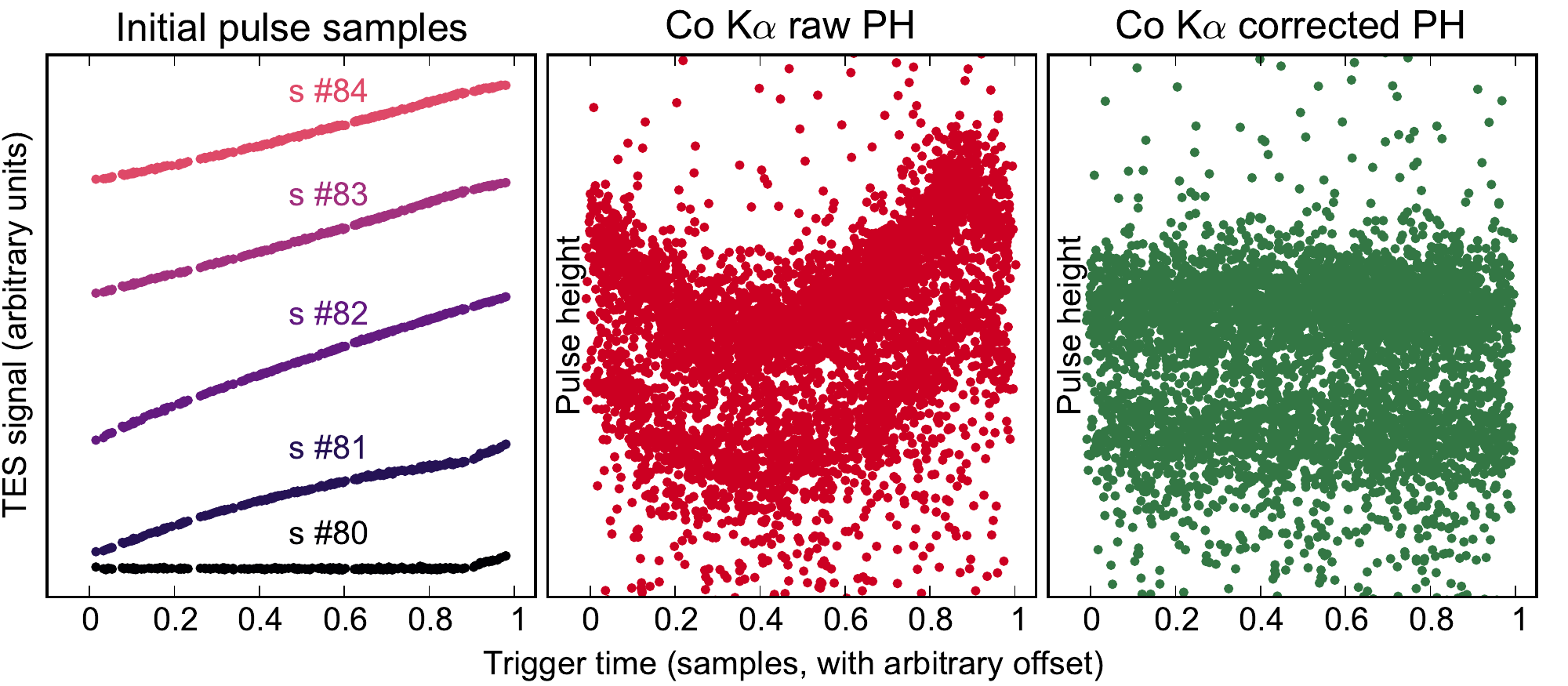}
\caption{ \label{fig:arrival_time}
The arrival-time systematic effect.
\emph{Left:} The primary cause is that pulses exceed the linear dynamic range of the SQUID readout system, at least for the first few samples on the pulse rising edge. Here, five samples near the trigger time (\#80--\#84) are shown for 300 Co K$\alpha$ pulses, as a function of their exact trigger-minus-arrival time (a larger trigger time $T$ indicates a pulse that triggered later, relative to its arrival time). The structure seen at small scales violates the assumptions of optimal filtering.  \emph{Center:} The filtered pulse heights for Co K$\alpha$ photons depend so strongly on $T$ that the systematic error of 5--10\,eV dominates the statistical resolution. \emph{Right:} The same pulses after a global correction chosen to align features across all $T$ and all energies, achieving a 4.3\,eV FWHM resolution. (Color figure online.)
}
\end{figure}

This arrival-time (AT) bias (Fig.\,\ref{fig:arrival_time}) indicates a failure of the simple linear model on which optimal filtering is premised. The failure is most pronounced at high frequencies, so one way to reduce the bias is by smoothing the optimal filter of Eq.~\ref{eq:opt_filt}. Such smoothing, of course, produces sub-optimal filtering and comes with a cost in signal-to-noise. We have found that smoothing by a 1-pole filter with 3\,dB point of 3\,kHz to 10\,kHz can be effective in reducing the AT bias level with no appreciable loss in energy resolution.  Like nearly all the approaches we have tried for mitigating the AT bias, the process of smoothing the optimal filter has two undesirable properties: (1)  it contains a free parameter which we have no principled way to select, and (2) it works very well in some measurements but hardly at all in others. Overall, though, it is a useful tool.

Some other promising ideas that fail to be universally beneficial include:
\begin{enumerate}
\item {\bf A 5-lag correlation.} \label{item:5lag} By constructing the filter (Eq.~\ref{eq:opt_filt}) with a model \mat{M} four samples shorter than our triggered pulse records, we can correlate the data with the filter at five different lags. Fitting a parabola to the values, we can find the lag at which it peaks (a very good estimate of the pulse arrival time) and the peak level. The latter can be somewhat less biased than the basic correlation value at lag-0, though this is not always so.
\item {\bf First-order time shift.} \label{item:dpdt} We can expand the pulse model to linear order in the arrival time shift $T$;  the leading term in $T$ is the time-derivative of the pulse model. Estimating this term from finite differences in the average pulse model, we can add this as a new column in the pulse model matrix \mat{M} and render the filtered pulse height insensitive  \cite{Fowler:2015} to the first-order effect of variations in $T$. 
\item {\bf Ignore the first samples in a pulse.} \label{item:gap} The first few samples are responsible for much of the AT bias. We can de-weight these samples by scaling down the corresponding columns and rows in \inv{R}, but this never fully removes the bias unless taken to the extreme such that it also fully removes the signal. This de-weighting also requires numerous parameters to be chosen without guidance.
\item {\bf Implement a more perfect trigger.} \label{item:shift} Most pulses satisfy the triggering condition in the sample immediately after a photon is absorbed in the sensor, but some will trigger only on the second. This latter group can be identified readily and shifted in time by 1 sample, effectively reducing the trigger threshold to zero.  While this step tends to remove a malign ``kink'' from the AT bias curves (see Fig.~\ref{fig:arrival_time} near $T=0.9$), it does not altogether eliminate the bias.
\item{\bf Sort pulses into $M$ arrival-time groups and filter separately.} Pulses can be ordered from early to late arrival times, and these can be partitioned into $M\sim10$ groups, with separate pulse models and optimal filters made for each group. This can work well, but the net result of the poorly known relative normalization of the $M$ filters is precisely to introduce a new form of AT bias.
\end{enumerate}
Almost all of these approaches work well at a single energy, and rather less well if we must apply them over a range of energies, a range that can often be 1000 times as large as the sensor resolution.


In careful studies at 4\,keV to 8\,keV, over a variety of TES designs and configurations, we have found no single combination of mitigation steps that universally eliminates the AT bias to the point of preserving the full intrinsic TES resolution.  Our usual practice in the past has been to smooth the optimal filter with a 1-pole low-pass filter having its 3\,dB point near 4\,kHz and to take the pulse height as the peak of the parabola fit to a 5-lag correlation function (item \#\ref{item:5lag}). The survey suggests that a simpler and equally effective plan is to use the zero-threshold trigger (\#\ref{item:shift}) and filters insensitive to the leading-order arrival time expansion term (or terms, \#\ref{item:dpdt}), and to take a single-lag correlation (thus, omit \#\ref{item:5lag}).

So long as we  push the sensor count to the highest numbers supported by the SQUID readout system, some amount of arrival-time bias appears inevitable. For the most sensitive work, an arrival-time correction akin to the one that reduces gain drift by sharpening up the output spectrum (Section~\ref{sec:gain_drift}) is necessary. Because the AT bias depends also on pulse energy, however, the correction cannot be as simple as the single number $\alpha$ that characterizes the gain drift. We are still developing best practices for sharpening spectra by correcting the AT bias with a method similar to the minimum-entropy method across a range of energies.

\sectionC{Future directions and conclusion}

The analysis of photon pulses described here works at a single photon energy or over a range of energies, though the latter case is made quite challenging because we adopt the wrong assumption that pulses (and their arrival-time systematics) are the same shape at all energies. We are actively pursuing methods that drop the single-shape assumption in favor of less restrictive pulse models. For example, we can project each pulse into a low-dimensional linear subspace based on the leading singular vectors in a training sample. If the data are first treated with a linear noise-whitening transformation \cite{Fowler:2015}, then this projection preserves the noise-weighting advantages of optimal filtering.
Projection into a linear subspace naturally accommodates both nonlinearity in energy and arrival-time effects, but it also presents a serious new challenge: energy calibration (without AT bias) of any point in the multi-dimensional manifold that the pulses span. 

We are also exploring ways to perform as much analysis as possible in real time, as data are acquired, without requiring human intervention to select and analyze training data (e.g., to make cuts or to find the optimal filter or a calibration curve). How best to do this is an open question, but it is becoming an important one as experiments are operated more hours per year with ever larger sensor arrays.

The analysis of x-ray microcalorimeter data to achieve maximum energy resolution and accuracy, while making efficient use of both computational and human resources, is a challenge with many faces. We have described here the basic analysis steps that we find to be important in nearly all spectrometer measurements and consider to be our current standard technique, along with some of the steps that we have analyzed and found lacking. We have not addressed the problem of the nonlinear conversion of pulse heights to energies, which is the subject of a future paper now in preparation.
Although the state of the art continues to evolve, we hope that this description of our standard approach can be useful to others operating spectrometers based on TES or similar microcalorimeters.

\vspace{-2mm}

\begin{acknowledgements}
The NIST Innovations in Measurement Science and the NASA Strategic Astrophysics Technology programs supported this work. JWF was  supported by an American Recovery and Reinvestment Act  fellowship. We thank Harvey Moseley, Simon Bandler, and Dale Fixsen for many discussions of pulse analysis and two helpful, anonymous referees.
\end{acknowledgements}

\end{document}